%% file: article.tex
\documentclass[showpacs,amsmath,amssymb,12pt]{revtex4}

\usepackage{graphicx}
\usepackage{dcolumn}
\usepackage{bm}
\usepackage{subfigure}

\begin{document}

\title{Numerical Bianchi I solutions in semi-classical gravitation}

\author{Sandro D. P. Vitenti}
\email{vitenti@isoftware.com.br}
\affiliation{Instituto de F\'isica, UnB \\ 
Campus Universit\'ario Darcy Ribeiro\\ 
Cxp 04455, 70919-970, Bras\'ilia DF \\
Brasil}
\author{Daniel M\"uller}
\email{muller@fis.unb.br}
\affiliation{Instituto de F\'isica, UnB \\ 
Campus Universit\'ario Darcy Ribeiro\\ 
Cxp 04455, 70919-970, Brasília DF \\
Brasil}

\date{\today}

\begin{abstract}
It is believed that soon after the Planck era, spacetime should have a semi-classical nature. In this 
context we consider quantum fields propagating in a classical gravitational field and study the 
backreaction of these fields, using the expected value of the energy-momentum tensor as source of 
the gravitational field. According to this theory, the escape from General Relativity theory is 
unavoidable. Two geometric counter-term are needed to regularize the divergences which come from 
the expected value. There is a parameter associated to each counter-term and in this work we found 
numerical solutions of this theory to particular initial conditions, for general Bianchi Type I spaces. We 
show that even though there are spurious solutions some of them can be considered physical. These 
physical solutions include de Sitter and Minkowski that are obtained asymptotically.

\end{abstract}
                            
\maketitle

\section{Introduction} 
The semi-classical theory consider the backreaction of quantum fields in a classical geometric 
background. It began about forty years ago with the introduction of the quadratic terms in the 
Riemann tensor \cite{DeWitt1} made by the renormalization of the energy-momentum tensor, and 
continued with a more complete work made by De Witt \cite{DeWitt2}, and since then, its 
consequences and applications are still under research, see for example \cite{Hu}. Some previous 
work \cite{huparker1} and \cite{huparker2}, studied the effect of particle creation in a Bianchi Type I 
spaces, and the renormalized coefficients of some of the higher
derivative terms obtained by renromalization were set to $0$.

Different of the usual Einstein-Hilbert action, the predicted  gravitational action allows differential 
equations with fourth order derivatives, which is called the full theory \cite{DeWitt1}, see also 
\cite{liv}. 

There are several problems connected to the full theory which prove that it is not consistent with our 
expectation of the present day physical world. Instability of flat space, Planck scale tidal forces, 
tachyonic propagation of gravitational particles, and violation of the positive energy theorem,  see for 
example \cite{prob}. Some of these results are obtained through linearization.  

In the very interesting articles \cite{Simon}, Simon and Parker-Simon, suggest that the theory should 
be seen in a perturbative fashion. Their method results in differential equations for the metric to only 
second order. The order reduction allows for the elimination of the non physical effects. For instance 
in absence of a classical energy momentum source, Einstein's General Relativity is identically 
recovered in \cite{Simon}. The above mentioned problems with the full theory are removed. The 
order reduction formalism should be the correct physical description in which the higher order terms 
are seen as a very small  perturbative correction. 

The full higher order theory was previously studied by Starobinsky \cite{S}, and more recently, also 
by Shapiro, Pelinson and others \cite{Shapiro}. Starobinsky idea was that the higher order terms 
could mimic a cosmological constant. In \cite{Shapiro} only the homogeneous and isotropic space 
time is studied. 

The full theory
with four time derivatives is addressed, which apparently was first
investigated in Tomita's article \cite{berkin} for general Bianchi I
spaces. They found that the presence of anisotropy contributes to the 
formation of the singularity. Berkin's work shows that a quadratic
Weyl theory is less stable than a quadratic Riemann scalar $R^2$.
Barrow and Hervik found exact and analytic solutions for anisotropic
quadratic gravity with $\Lambda \neq 0$ that do not approach a de Sitter
space time. In that article Barrow and Hervik discuss Bianchi
types $II$ and $VI_h$ and also a very interesting stability criterion concerning 
small anisotropies. There is also a recent article by Clifton and Barrow in 
which Kasner type solutions are addressed. H. J. Schmidt does a recent and very interesting  
review of higher order gravity theories in connection to cosmology \cite{hjs}. 

In this present work, only the vacuum energy momentum classical source is 
considered in the higher derivative theory. It should be valid 
soon after the Planck era neglecting any particle creation that took place at that time. The full 
theory predicts explosions and formation of physical singularities depending in the parameters and 
initial conditions. From this point of view the full theory is
certainly not a complete theory. In fact, the 
exact numerical solutions for Bianchi I spaces seem to reproduce all the problems mentioned above, 
this question was not investigated in this present work. 

But for some values of the parameters and very particular initial conditions, physically consistent 
solutions are obtained. The isotropization of spacetime also occurs with zero cosmological constant. 
The scalar Riemann four curvature oscillates near a constant value with decreasing amplitude. 

The following conventions and unit choice are taken $R^a_{bcd}=\Gamma^a_{bd,c}-...$, 
$R_{ab}=R^c_{acb}$, $R=R^a_a$, metric signature $-+++$, Latin symbols run from $0-3$, Greek 
symbols run from $1-3$ and $G=\hbar=c=1$.  

\section{Full theory and numerical solutions}

The Lagrange function is, 
\begin{equation}
{\cal L}=\sqrt{-g}\left[\Lambda + R+
\alpha\left( R^{ab}R_{ab}-\frac{1}{3}R^2\right)+\beta R^2\right] {\label{acao}} + {\cal L}_c\,.
\end{equation}
Metric variations in the above action results in
\begin{equation}
E_{ab}=G_{ab}+\left(\beta-\frac{1}{3}\alpha\right)H^{(1)}_{ab}+\alpha H^{(2)}_{ab}
-T_{ab}-\frac{1}{2}\Lambda g_{ab},
\label{eqtotal}
\end{equation}
where
\begin{eqnarray}
&&H^{(1)}_{ab}=\frac{1}{2}g_{ab}R^2-2RR_{ab}+2R_{;ab}-2\square Rg_{ab},\\
&&H^{(2)}_{ab}=\frac{1}{2}g_{ab}R_{mn}R^{mn}+R_{;ab}-2R^{cn}R_{cbna}-\square R_{ab}-\frac{1}{2}\square Rg_{ab}, \\
&&G_{ab}=R_{ab}-\frac{1}{2}g_{ab}R,
\end{eqnarray}
and $T_{ab}$ is the energy momentum source, which comes from the classical part of the 
Lagrangian ${\cal L}_c$. Only classical vacuum solutions $T_{ab}={\cal L}_c=0$  will be 
considered in this section since it seems the most natural condition soon after the Planck era.
 
The covariant divergence of the above tensors are identically zero due their variational definition. The 
following Bianchi Type I line element is considered
\begin{equation}
ds^2=-dt^2+e^{2a_1(t)}dx^2+e^{2a_2(t)}dy^2+e^{2a_3(t)}dz^2 \label{elinha},
\end{equation}
which is a general spatially flat and anisotropic space, with proper time $t$. With this line element all 
the tensors which enter the expressions are  diagonal. The substitution of \eqref{elinha} in 
\eqref{eqtotal} with $T_{ab}=0$, results for the spatial part of \eqref{eqtotal}, in differential 
equations of the type 
\begin{eqnarray}
&&\frac{d^{4}}{dt^4}a_1=f_1\left( \frac{d^3}{dt^3}a_i,\ddot{a}_i,\dot{a}_i
\right) \label{edo1}\\
&&\frac{d^{4}}{dt^4}a_2=f_2\left( \frac{d^3}{dt^3}a_i,\ddot{a}_i,\dot{a}_i
\right)\\
&&\frac{d^{4}}{dt^4}a_3=f_3\left( \frac{d^3}{dt^3}a_i,\ddot{a}_i,\dot{a}_i
\right)\label{edo3},
\end{eqnarray}
where the functions $f_i$ involve the derivatives of $a_1,\;a_2,\;a_3$, in a polynomial fashion. The 
very interesting article \cite{Noakes} shows that the theory which follows from \eqref{acao} has a 
well posed initial value problem. In \cite{Noakes} the differential equations for the metric are written in 
a form suitable for the application of the theorem of Leray \cite{Leray}.

Instead of going through the general construction given in \cite{Noakes}, in this particular case, the 
existence and uniqueness of the solutions of \eqref{edo1}-\eqref{edo3} are more simply understood 
in \cite{reedsimon}. 

Besides the equations \eqref{edo1}-\eqref{edo3}, we have the temporal part of \eqref{eqtotal}. To 
understand the role of this equation we have first to study the covariant divergence of the equation 
\eqref{eqtotal}, $$\nabla_aE^{ab} = \partial_aE^{ab} + \Gamma^a_{ac}E^{cb} + 
\Gamma^{b}_{ac} E^{ac} = 0,$$ but since we are using \eqref{edo1}-\eqref{edo3} to integrate 
the system numerically $E^{ii}\equiv 0,$ then 
\begin{equation}
\partial_0E^{00} + \Gamma^a_{a0} E^{00} + 
\Gamma^{0}_{00} E^{00} = 0.
\label{vinculo}
\end{equation}
Therefore if $E_{00}=0$ initially, it will remain zero at any instant. 
Therefore the equation $E_{00}$ acts as a constraint on the initial conditions and we use it to test the 
accuracy of our results.

For a space-like vector $v^\alpha$ and a time-like vector $t^a=(1,0,0,0)$ tidal forces are given by 
the geodesic deviation equations
\begin{eqnarray*}
&&t^a\nabla_a v^\alpha=R^\alpha_{mn\beta}t^mt^nv^\beta \\
&&t^a\nabla_a v^\alpha=R^\alpha_{00\beta}v^{\beta}\\
&&R^\alpha_{00\beta}=\delta_{\alpha\beta}\left(\dot{a}_\alpha^2
+\ddot{a}_\alpha\right).
\end{eqnarray*}
The theory predicted by \eqref{acao} is believed to be correct if the tidal forces are less than $1$ in 
Plank units units, 
\begin{equation}
|R^\alpha_{0\alpha0}|\leq 1\;\; \text{(no summation.)}\label{mare}
\end{equation}
When this condition is not satisfied, quantum effects could introduce further modifications into 
\eqref{acao}.

In order to integrate these equations we used a open source and well tested C library, the GSL 
\emph{GNU Scientific Library} \cite{gsl}, we used several algorithms provided by this library like 
\emph{Embedded Runge-Kutta-Fehlberg (4, 5) method} and \emph{Implicit Bulirsch-Stoer method of 
Bader and Deuflhard}, we also used \emph{Maple} to calculate the equations and a \emph{Perl} 
script developed by us to generate the C source code from the Maple output, and finally we used a 
\emph{AMD Athlon(TM) XP 2600+} to integrate the equations.
 
\subsection{Numerical Solutions}
For particular initial conditions and values for the parameters $\alpha, \,\beta,\,\Lambda$  
consistent numerical solutions for \eqref{eqtotal} are shown in 
FIG. \ref{fig:con0} and \ref{fig:con1}. In FIG. \ref{fig:con0} 
$\alpha=0.1,\;\beta=-0.1,\;\Lambda=0$ and in FIG. \ref{fig:con1} $\alpha=1000,\;\beta=-1000,\;\Lambda=0$

\input{fig_full}

The only non null initial conditions are chosen 
\begin{center}
\begin{tabular}{|c|c|c|c|}
\hline $\dot{a}_1(0)$ & $\dot{a}_2(0)$  & $\dot{a}_3(0)$ & $d^3a_1/dt^3$ \\ 
\hline $1\times 10^{-1}$ & $2\times 10^{-1}$  & $7\times 10^{-1}$  & $-4.3248\times 10^{-1}$\\ 
\hline 
\end{tabular} 
\end{center}
for FIG \ref{fig:con0} and
\begin{center}
\begin{tabular}{|c|c|c|c|}
\hline $\dot{a}_1(0)$ & $\dot{a}_2(0)$  & $\dot{a}_3(0)$ & $d^3a_1/dt^3$  \\ 
\hline $1\times 10^{-1}$ & $2\times 10^{-1}$  & $7\times 10^{-1}$ & $1.1076\times 10^{-1}$  \\ 
\hline 
\end{tabular} 
\end{center}
for FIG.\ref{fig:con1}, such that the $00$ component of \eqref{eqtotal} vanishes initially. 
According to \eqref{vinculo}, it is numerically checked that $|E_{00}|<10^{-16}$ for the time interval 
in FIG. \ref{fig:con0} and  $|E_{00}|<10^{-12}$ for FIG. \ref{fig:con1}.

The condition given in \eqref{mare} is satisfied initially, and during the time evolution in 
the solutions plotted in FIG. 
\ref{fig:con0}-\ref{fig:con1}. Certainly these are physically accepted solutions since 
FIG. \ref{fig:con0} gives Minkowski space and FIG. \ref{fig:con1} gives de Sitter space asymptotically.
\subsection{Explosions} 
In FIG. \ref{fig:ex2} the values of the parameters are different 
\[
\alpha=-0.1, \, \beta=0.1, \, \Lambda=0.
\] 

And again the only non null initial conditions are 
\begin{center}
\begin{tabular}{|c|c|c|c|}
\hline $\dot{a}_1(0)$ & $\dot{a}_2(0)$  & $\dot{a}_3(0)$ & $d^3a_1/dt^3$  \\ 
\hline $1\times 10^{-1}$ & $2\times 10^{-1}$  & $7\times 10^{-1}$ & $6.5412\times 10^{-1}$  \\ 
\hline 
\end{tabular} 
\end{center}
are such that the $00$ component of \eqref{eqtotal} vanishes. According to \eqref{vinculo}, it is 
numerically checked that $|E_{00}|<10^{-14}$ for the time interval in FIG. \ref{fig:ex2}.

It can be seen that although the initial conditions satisfy \eqref{mare}, $|R^\alpha_{0\alpha0}|$ (no 
summation), assumes arbitrary large increasing positive values, which indicates the existence of 
explosions. In FIG. \ref{fig:ex2}, it is shown that the curvature scalar increases to arbitrary large 
positive values.  
\subsection{Formation of singularities}
The following values for 
\[
\alpha=0.1, \, \beta=0.1, \, \Lambda=0, 
\] 
are the same for FIG. \ref{fig:ex1} and FIG. \ref{fig:ex3}. In FIG. \ref{fig:ex1}
the non null initial conditions are
\begin{center}
\begin{tabular}{|c|c|c|c|}
\hline $\dot{a}_1(0)$ & $\dot{a}_2(0)$  & $\dot{a}_3(0)$ & $d^3a_1/dt^3$  \\ 
\hline $1\times 10^{-1}$ & $2\times 10^{-1}$  & $7\times 10^{-1}$ & $7.3955\times 10^{-1}$  \\ 
\hline 
\end{tabular} 
\end{center}
and for FIG. \ref{fig:ex3}
\begin{center}
\begin{tabular}{|c|c|c|c|}
\hline $\dot{a}_1(0)$ & $\dot{a}_2(0)$  & $\dot{a}_3(0)$ & $d^3a_1/dt^3$  \\ 
\hline $1\times 10^{-1}$ & $2\times 10^{-1}$  & $-2\times 10^{-1}$ & $-8.1350\times 10^{-1}$  \\ 
\hline 
\end{tabular} 
\end{center}
such that the $00$ component of \eqref{eqtotal} vanishes. According to \eqref{vinculo}, it is 
numerically checked that $|E_{00}|<10^{-10}$ for the time interval in 
FIG. \ref{fig:ex1} and FIG. \ref{fig:ex3}. 
Since both the scalar curvature tensor $R$ and the squared Ricci tensor $R_{ab}R^{ab}$, 
increase abruptly, the solutions shown in FIG. \ref{fig:ex1} and FIG. \ref{fig:ex3} 
are understood as singular type.
The numerical error increases very fast when the solution comes closer to the singularity, which is 
expected.
\section{Conclusions}
In the present work it is considered general anisotropic Bianchi I homogeneous spacetimes. The full 
theory with four time derivatives problem are addressed. 

Only the vacuum energy momentum classical source is considered in the full theory. It should be valid 
soon after the Planck era and vacuum classical source seems the most natural condition. The full 
theory predicts explosions and formation of physical singularities depending in the parameters and 
initial conditions. From this point of view the full theory is certainly not a complete theory. In fact, the 
exact numerical solutions for Bianchi I spaces seem to reproduce all the problems mentioned above, 
this question was not investigated in this present work.
 
However, for some values of the parameters and very particular initial conditions, physically 
consistent solutions are obtained. The isotropization of space time also occurs with zero cosmological 
constant. The scalar Riemann four curvature oscillates near constant values with decreasing amplitude. 

The above is understood as follows. The formation of singularities and Plank type explosions indicate 
that the theory certainly is not complete. The existence of physical consistent solutions show that 
the theory could have a space-time region of validity.

Although we did not attempt 
a detailed verification, the
numerical solutions obtained in this present work show no
contradiction to the interesting previous calculations concerning anisotropies
\cite{berkin}. The analytical solutions found by Barrow-Hervik and
Clifton-Barrow can be understood as limit sets in the space of
solutions of the quadratic theory \eqref{acao}.

\begin{acknowledgements}
S. D. P. Vitenti wishes to thank the Brazilian agency CAPES for
financial support. D. M. wishes to thank the Brazilian project {\it
Nova F\'\i sica no Espa\c co}. We shall like to thank the anonymous
referees, for improvements.
\end{acknowledgements} 

\end{document}

%% file: fig_full.tex
\begin{figure}[htpb]
  \begin{center}
    \resizebox{70mm}{!}{\input{plot_full_1}}
    \resizebox{70mm}{!}{\input{plot_full_v1}}
  \end{center}
  \begin{center}
    \resizebox{70mm}{!}{\input{plot_full_R1}}
    \resizebox{70mm}{!}{\input{plot_full_Rc21}}
  \end{center}
  \caption{Using $\alpha=0.1,\,\beta=-0.1,\,\Lambda=0$}
  \label{fig:con0}
\end{figure}
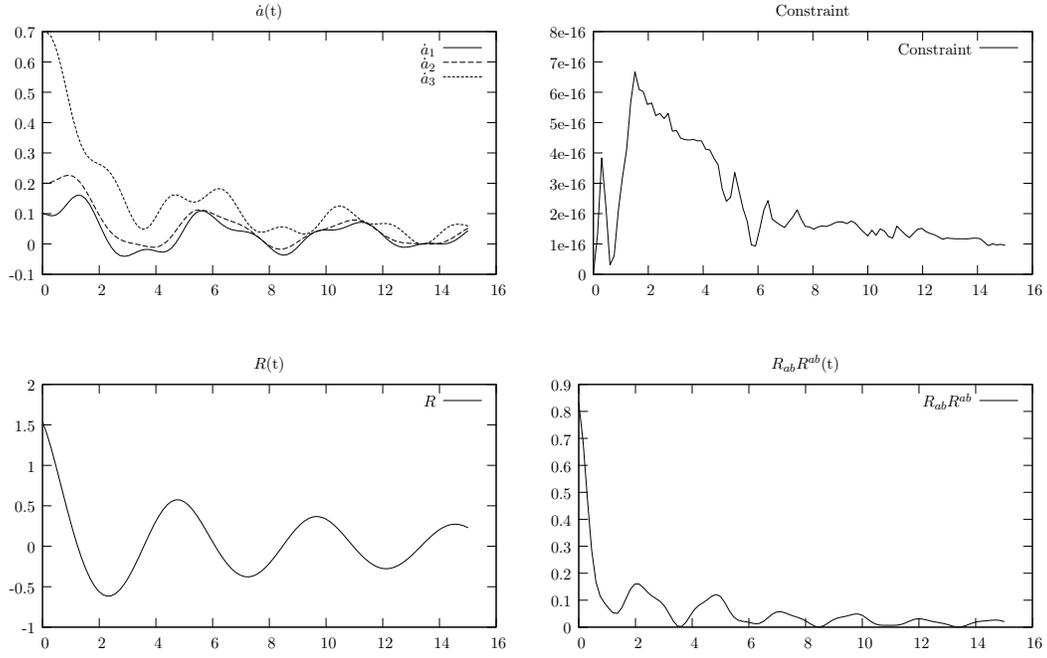

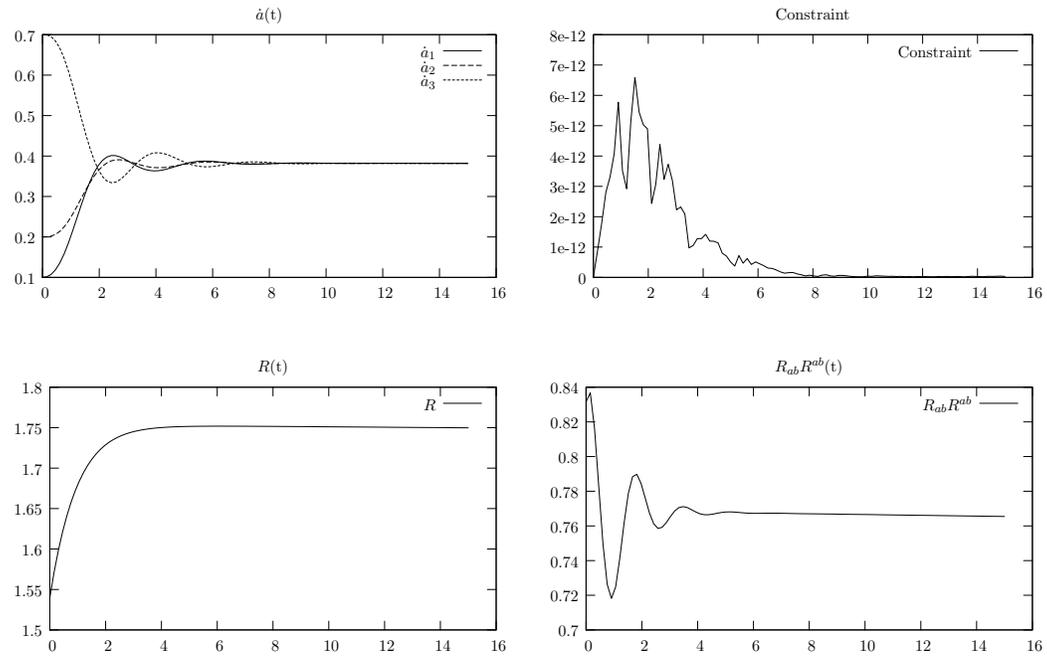
\begin{figure}[htpb]
  \begin{center}
    \resizebox{70mm}{!}{\input{plot_full_11}}
    \resizebox{70mm}{!}{\input{plot_full_v11}}
  \end{center}
  \begin{center}
    \resizebox{70mm}{!}{\input{plot_full_R11}}
    \resizebox{70mm}{!}{\input{plot_full_Rc211}}
  \end{center}
  \caption{Using $\alpha=1000,\,\beta=-1000,\,\Lambda=0$}
  \label{fig:con1}
\end{figure}

\begin{figure}[htpb]
  \begin{center}
    \resizebox{70mm}{!}{\input{plot_full_0}}
    \resizebox{70mm}{!}{\input{plot_full_v0}}
  \end{center}
  \begin{center}
    \resizebox{70mm}{!}{\input{plot_full_R0}}
    \resizebox{70mm}{!}{\input{plot_full_Rc20}}
  \end{center}
  \caption{Using $\alpha=0.1,\,\beta=0.1,\,\Lambda=0$}
  \label{fig:ex1}
\end{figure}

\begin{figure}[htpb]
  \begin{center}
    \resizebox{70mm}{!}{\input{plot_full_4}}
    \resizebox{70mm}{!}{\input{plot_full_v4}}
  \end{center}
  \begin{center}
    \resizebox{70mm}{!}{\input{plot_full_R4}}
    \resizebox{70mm}{!}{\input{plot_full_Rc24}}
  \end{center}
  \caption{Using $\alpha=-0.1,\,\beta=0.1,\,\Lambda=0$}
  \label{fig:ex2}
\end{figure}

\begin{figure}[htpb]
  \begin{center}
    \resizebox{70mm}{!}{\input{plot_full_16}}
    \resizebox{70mm}{!}{\input{plot_full_v16}}
  \end{center}
  \begin{center}
    \resizebox{70mm}{!}{\input{plot_full_R16}}
    \resizebox{70mm}{!}{\input{plot_full_Rc216}}
  \end{center}
  \caption{Using $\alpha=0.1,\,\beta=0.1,\,\Lambda=0$}
  \label{fig:ex3}
\end{figure}

%% file: plot_full_1.tex
\begin{picture}(0,0)%
\includegraphics{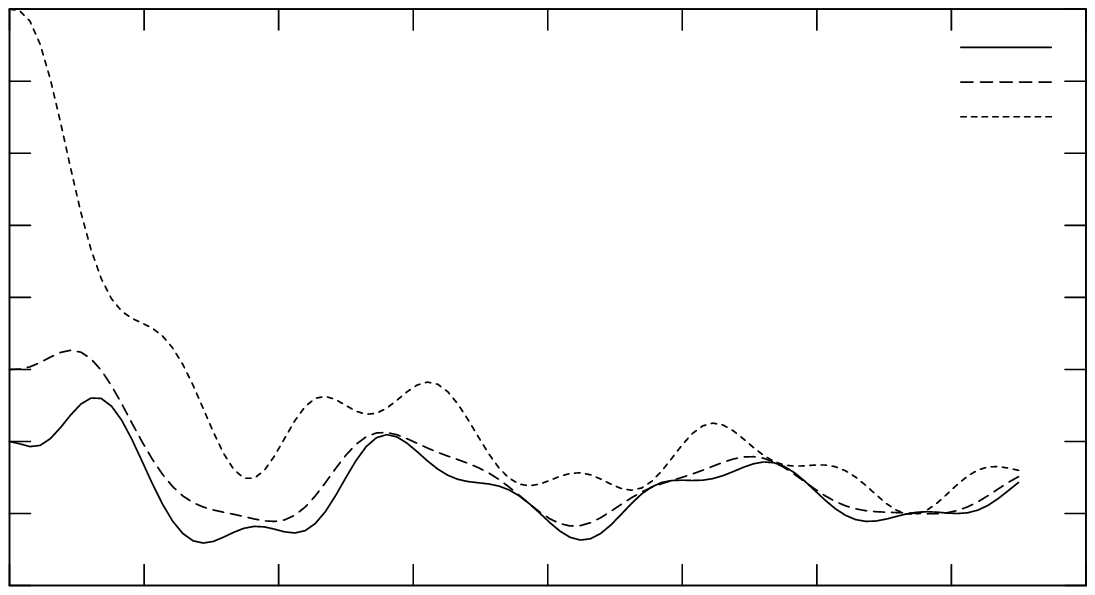}%
\end{picture}%
\begingroup
\setlength{\unitlength}{0.0200bp}%
\begin{picture}(18000,10800)(0,0)%
\put(1500,1000){\makebox(0,0)[r]{\strut{}-0.1}}%
\put(1500,2037){\makebox(0,0)[r]{\strut{} 0}}%
\put(1500,3075){\makebox(0,0)[r]{\strut{} 0.1}}%
\put(1500,4112){\makebox(0,0)[r]{\strut{} 0.2}}%
\put(1500,5150){\makebox(0,0)[r]{\strut{} 0.3}}%
\put(1500,6187){\makebox(0,0)[r]{\strut{} 0.4}}%
\put(1500,7225){\makebox(0,0)[r]{\strut{} 0.5}}%
\put(1500,8262){\makebox(0,0)[r]{\strut{} 0.6}}%
\put(1500,9300){\makebox(0,0)[r]{\strut{} 0.7}}%
\put(1750,500){\makebox(0,0){\strut{} 0}}%
\put(3688,500){\makebox(0,0){\strut{} 2}}%
\put(5625,500){\makebox(0,0){\strut{} 4}}%
\put(7563,500){\makebox(0,0){\strut{} 6}}%
\put(9500,500){\makebox(0,0){\strut{} 8}}%
\put(11438,500){\makebox(0,0){\strut{} 10}}%
\put(13375,500){\makebox(0,0){\strut{} 12}}%
\put(15313,500){\makebox(0,0){\strut{} 14}}%
\put(17250,500){\makebox(0,0){\strut{} 16}}%
\put(9500,10050){\makebox(0,0){\strut{}$\dot{a}$(t)}}%
\put(15200,8750){\makebox(0,0)[r]{\strut{}$\dot{a}_1$}}%
\put(15200,8250){\makebox(0,0)[r]{\strut{}$\dot{a}_2$}}%
\put(15200,7750){\makebox(0,0)[r]{\strut{}$\dot{a}_3$}}%
\end{picture}%
\endgroup
 

%% file: plot_full_v1.tex
\begin{picture}(0,0)%
\includegraphics{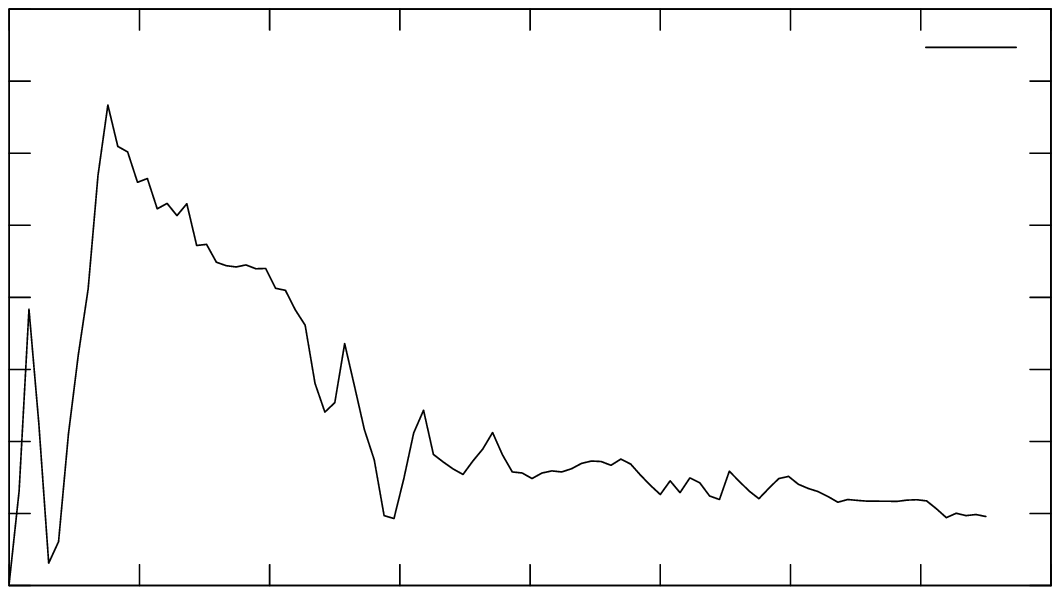}%
\end{picture}%
\begingroup
\setlength{\unitlength}{0.0200bp}%
\begin{picture}(18000,10800)(0,0)%
\put(2000,1000){\makebox(0,0)[r]{\strut{} 0}}%
\put(2000,2037){\makebox(0,0)[r]{\strut{} 1e-16}}%
\put(2000,3075){\makebox(0,0)[r]{\strut{} 2e-16}}%
\put(2000,4112){\makebox(0,0)[r]{\strut{} 3e-16}}%
\put(2000,5150){\makebox(0,0)[r]{\strut{} 4e-16}}%
\put(2000,6188){\makebox(0,0)[r]{\strut{} 5e-16}}%
\put(2000,7225){\makebox(0,0)[r]{\strut{} 6e-16}}%
\put(2000,8263){\makebox(0,0)[r]{\strut{} 7e-16}}%
\put(2000,9300){\makebox(0,0)[r]{\strut{} 8e-16}}%
\put(2250,500){\makebox(0,0){\strut{} 0}}%
\put(4125,500){\makebox(0,0){\strut{} 2}}%
\put(6000,500){\makebox(0,0){\strut{} 4}}%
\put(7875,500){\makebox(0,0){\strut{} 6}}%
\put(9750,500){\makebox(0,0){\strut{} 8}}%
\put(11625,500){\makebox(0,0){\strut{} 10}}%
\put(13500,500){\makebox(0,0){\strut{} 12}}%
\put(15375,500){\makebox(0,0){\strut{} 14}}%
\put(17250,500){\makebox(0,0){\strut{} 16}}%
\put(9750,10050){\makebox(0,0){\strut{}Constraint}}%
\put(15200,8750){\makebox(0,0)[r]{\strut{}Constraint}}%
\end{picture}%
\endgroup
 

%% file: plot_full_R1.tex
\begin{picture}(0,0)%
\includegraphics{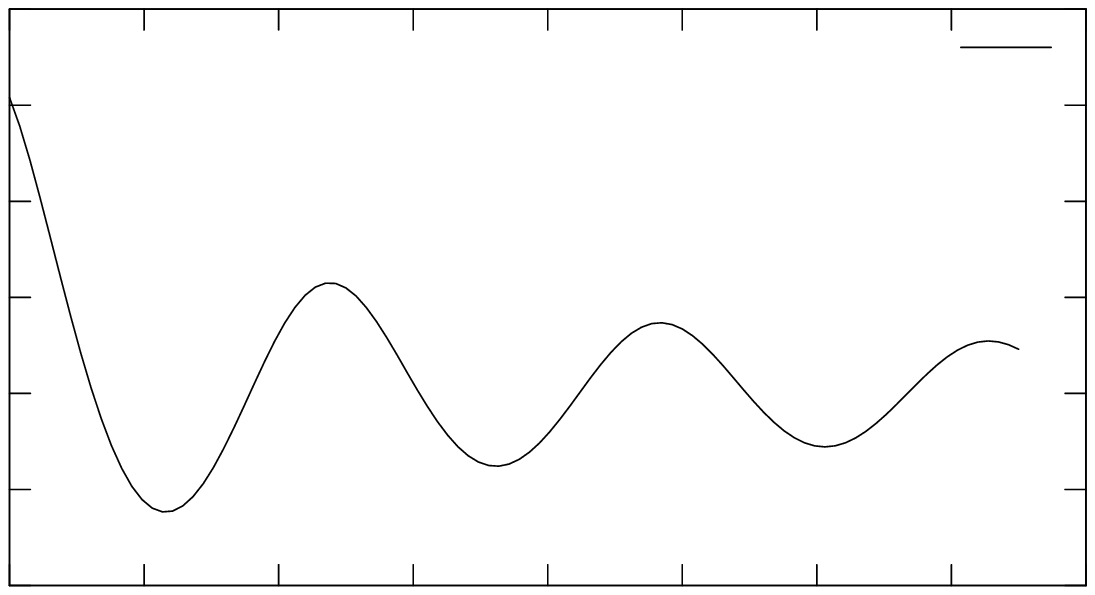}%
\end{picture}%
\begingroup
\setlength{\unitlength}{0.0200bp}%
\begin{picture}(18000,10800)(0,0)%
\put(1500,1000){\makebox(0,0)[r]{\strut{}-1}}%
\put(1500,2383){\makebox(0,0)[r]{\strut{}-0.5}}%
\put(1500,3767){\makebox(0,0)[r]{\strut{} 0}}%
\put(1500,5150){\makebox(0,0)[r]{\strut{} 0.5}}%
\put(1500,6533){\makebox(0,0)[r]{\strut{} 1}}%
\put(1500,7917){\makebox(0,0)[r]{\strut{} 1.5}}%
\put(1500,9300){\makebox(0,0)[r]{\strut{} 2}}%
\put(1750,500){\makebox(0,0){\strut{} 0}}%
\put(3688,500){\makebox(0,0){\strut{} 2}}%
\put(5625,500){\makebox(0,0){\strut{} 4}}%
\put(7563,500){\makebox(0,0){\strut{} 6}}%
\put(9500,500){\makebox(0,0){\strut{} 8}}%
\put(11438,500){\makebox(0,0){\strut{} 10}}%
\put(13375,500){\makebox(0,0){\strut{} 12}}%
\put(15313,500){\makebox(0,0){\strut{} 14}}%
\put(17250,500){\makebox(0,0){\strut{} 16}}%
\put(9500,10050){\makebox(0,0){\strut{}$R$(t)}}%
\put(15200,8750){\makebox(0,0)[r]{\strut{}$R$}}%
\end{picture}%
\endgroup
 

%% file: plot_full_Rc21.tex
\begin{picture}(0,0)%
\includegraphics{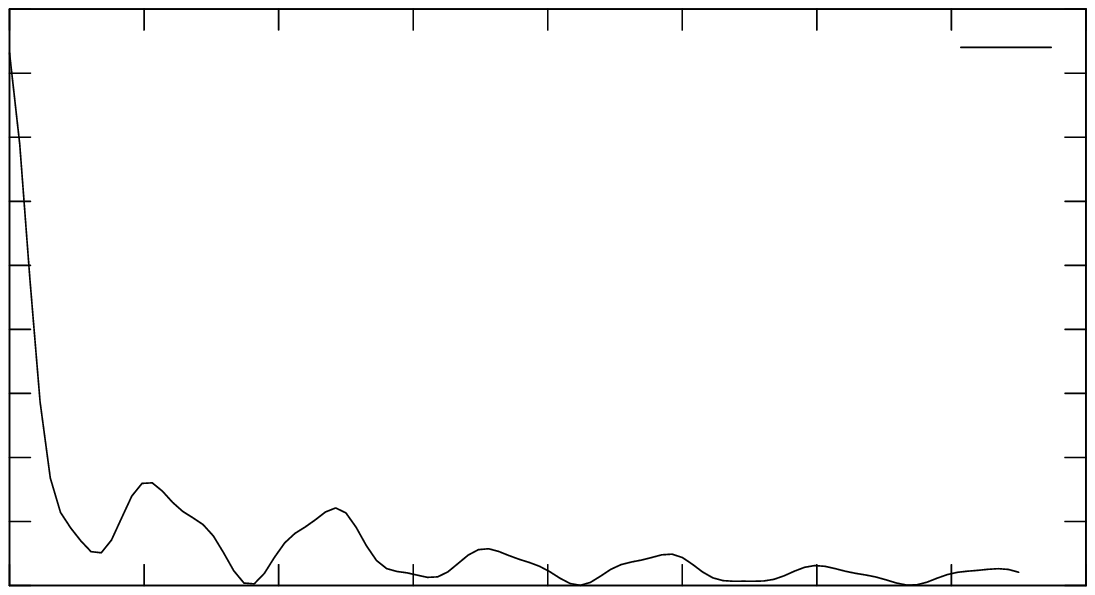}%
\end{picture}%
\begingroup
\setlength{\unitlength}{0.0200bp}%
\begin{picture}(18000,10800)(0,0)%
\put(1500,1000){\makebox(0,0)[r]{\strut{} 0}}%
\put(1500,1922){\makebox(0,0)[r]{\strut{} 0.1}}%
\put(1500,2844){\makebox(0,0)[r]{\strut{} 0.2}}%
\put(1500,3767){\makebox(0,0)[r]{\strut{} 0.3}}%
\put(1500,4689){\makebox(0,0)[r]{\strut{} 0.4}}%
\put(1500,5611){\makebox(0,0)[r]{\strut{} 0.5}}%
\put(1500,6533){\makebox(0,0)[r]{\strut{} 0.6}}%
\put(1500,7456){\makebox(0,0)[r]{\strut{} 0.7}}%
\put(1500,8378){\makebox(0,0)[r]{\strut{} 0.8}}%
\put(1500,9300){\makebox(0,0)[r]{\strut{} 0.9}}%
\put(1750,500){\makebox(0,0){\strut{} 0}}%
\put(3688,500){\makebox(0,0){\strut{} 2}}%
\put(5625,500){\makebox(0,0){\strut{} 4}}%
\put(7563,500){\makebox(0,0){\strut{} 6}}%
\put(9500,500){\makebox(0,0){\strut{} 8}}%
\put(11438,500){\makebox(0,0){\strut{} 10}}%
\put(13375,500){\makebox(0,0){\strut{} 12}}%
\put(15313,500){\makebox(0,0){\strut{} 14}}%
\put(17250,500){\makebox(0,0){\strut{} 16}}%
\put(9500,10050){\makebox(0,0){\strut{}$R_{ab}R^{ab}$(t)}}%
\put(15200,8750){\makebox(0,0)[r]{\strut{}$R_{ab}R^{ab}$}}%
\end{picture}%
\endgroup
 

%% file: plot_full_11.tex
\begin{picture}(0,0)%
\includegraphics{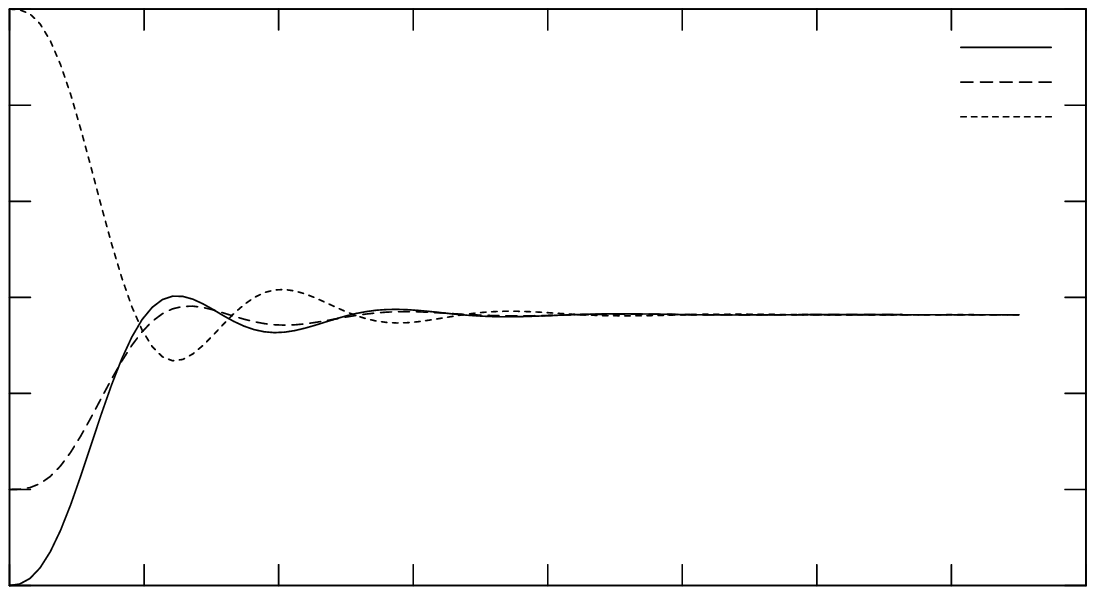}%
\end{picture}%
\begingroup
\setlength{\unitlength}{0.0200bp}%
\begin{picture}(18000,10800)(0,0)%
\put(1500,1000){\makebox(0,0)[r]{\strut{} 0.1}}%
\put(1500,2383){\makebox(0,0)[r]{\strut{} 0.2}}%
\put(1500,3767){\makebox(0,0)[r]{\strut{} 0.3}}%
\put(1500,5150){\makebox(0,0)[r]{\strut{} 0.4}}%
\put(1500,6533){\makebox(0,0)[r]{\strut{} 0.5}}%
\put(1500,7917){\makebox(0,0)[r]{\strut{} 0.6}}%
\put(1500,9300){\makebox(0,0)[r]{\strut{} 0.7}}%
\put(1750,500){\makebox(0,0){\strut{} 0}}%
\put(3688,500){\makebox(0,0){\strut{} 2}}%
\put(5625,500){\makebox(0,0){\strut{} 4}}%
\put(7563,500){\makebox(0,0){\strut{} 6}}%
\put(9500,500){\makebox(0,0){\strut{} 8}}%
\put(11438,500){\makebox(0,0){\strut{} 10}}%
\put(13375,500){\makebox(0,0){\strut{} 12}}%
\put(15313,500){\makebox(0,0){\strut{} 14}}%
\put(17250,500){\makebox(0,0){\strut{} 16}}%
\put(9500,10050){\makebox(0,0){\strut{}$\dot{a}$(t)}}%
\put(15200,8750){\makebox(0,0)[r]{\strut{}$\dot{a}_1$}}%
\put(15200,8250){\makebox(0,0)[r]{\strut{}$\dot{a}_2$}}%
\put(15200,7750){\makebox(0,0)[r]{\strut{}$\dot{a}_3$}}%
\end{picture}%
\endgroup
 

%% file: plot_full_v11.tex
\begin{picture}(0,0)%
\includegraphics{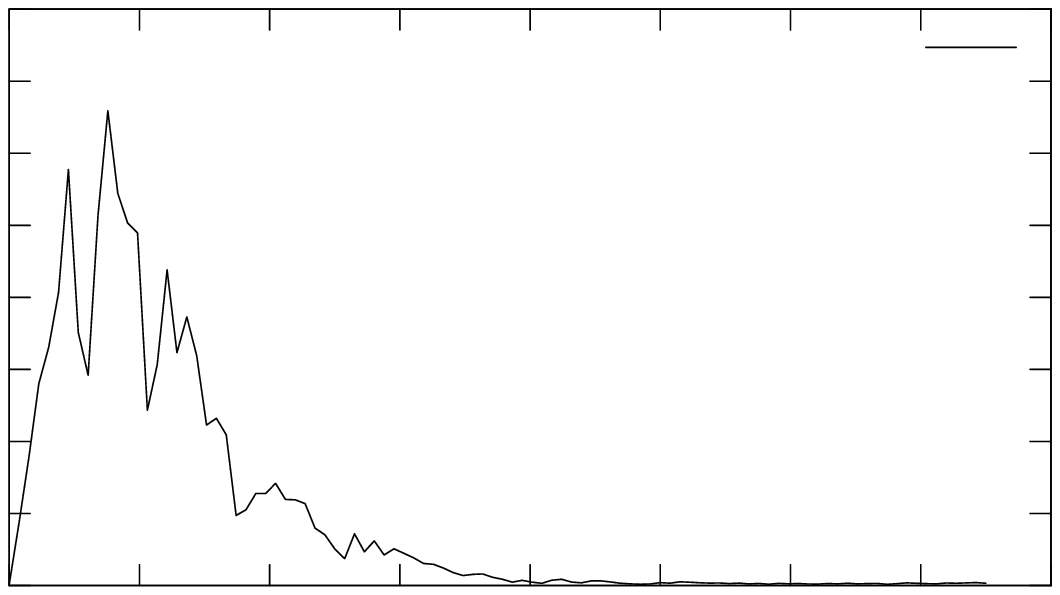}%
\end{picture}%
\begingroup
\setlength{\unitlength}{0.0200bp}%
\begin{picture}(18000,10800)(0,0)%
\put(2000,1000){\makebox(0,0)[r]{\strut{} 0}}%
\put(2000,2037){\makebox(0,0)[r]{\strut{} 1e-12}}%
\put(2000,3075){\makebox(0,0)[r]{\strut{} 2e-12}}%
\put(2000,4113){\makebox(0,0)[r]{\strut{} 3e-12}}%
\put(2000,5150){\makebox(0,0)[r]{\strut{} 4e-12}}%
\put(2000,6187){\makebox(0,0)[r]{\strut{} 5e-12}}%
\put(2000,7225){\makebox(0,0)[r]{\strut{} 6e-12}}%
\put(2000,8262){\makebox(0,0)[r]{\strut{} 7e-12}}%
\put(2000,9300){\makebox(0,0)[r]{\strut{} 8e-12}}%
\put(2250,500){\makebox(0,0){\strut{} 0}}%
\put(4125,500){\makebox(0,0){\strut{} 2}}%
\put(6000,500){\makebox(0,0){\strut{} 4}}%
\put(7875,500){\makebox(0,0){\strut{} 6}}%
\put(9750,500){\makebox(0,0){\strut{} 8}}%
\put(11625,500){\makebox(0,0){\strut{} 10}}%
\put(13500,500){\makebox(0,0){\strut{} 12}}%
\put(15375,500){\makebox(0,0){\strut{} 14}}%
\put(17250,500){\makebox(0,0){\strut{} 16}}%
\put(9750,10050){\makebox(0,0){\strut{}Constraint}}%
\put(15200,8750){\makebox(0,0)[r]{\strut{}Constraint}}%
\end{picture}%
\endgroup
 

%% file: plot_full_R11.tex
\begin{picture}(0,0)%
\includegraphics{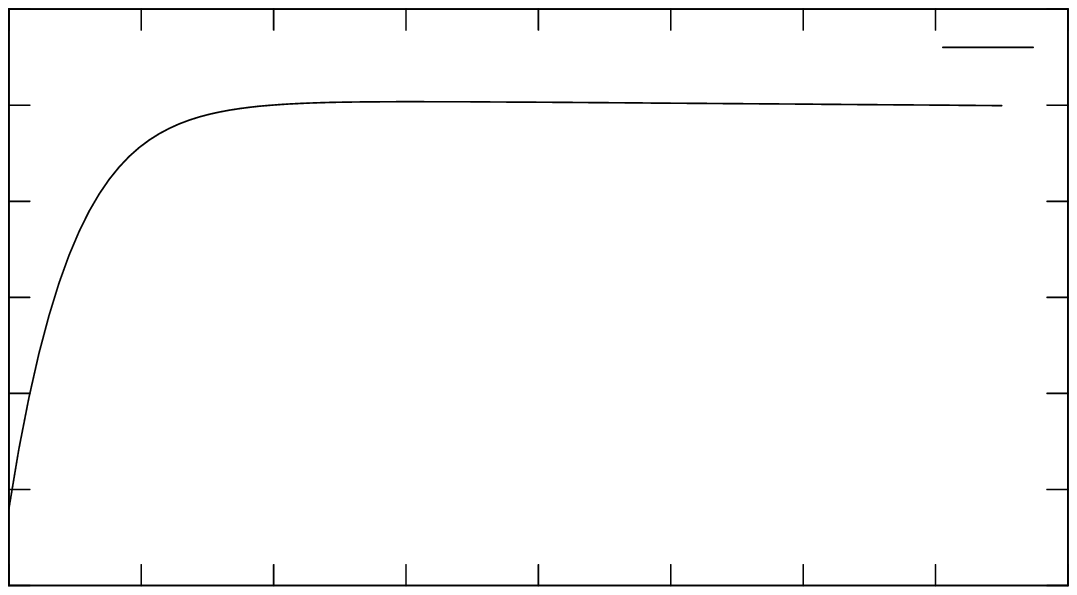}%
\end{picture}%
\begingroup
\setlength{\unitlength}{0.0200bp}%
\begin{picture}(18000,10800)(0,0)%
\put(1750,1000){\makebox(0,0)[r]{\strut{} 1.5}}%
\put(1750,2383){\makebox(0,0)[r]{\strut{} 1.55}}%
\put(1750,3767){\makebox(0,0)[r]{\strut{} 1.6}}%
\put(1750,5150){\makebox(0,0)[r]{\strut{} 1.65}}%
\put(1750,6533){\makebox(0,0)[r]{\strut{} 1.7}}%
\put(1750,7917){\makebox(0,0)[r]{\strut{} 1.75}}%
\put(1750,9300){\makebox(0,0)[r]{\strut{} 1.8}}%
\put(2000,500){\makebox(0,0){\strut{} 0}}%
\put(3906,500){\makebox(0,0){\strut{} 2}}%
\put(5813,500){\makebox(0,0){\strut{} 4}}%
\put(7719,500){\makebox(0,0){\strut{} 6}}%
\put(9625,500){\makebox(0,0){\strut{} 8}}%
\put(11531,500){\makebox(0,0){\strut{} 10}}%
\put(13438,500){\makebox(0,0){\strut{} 12}}%
\put(15344,500){\makebox(0,0){\strut{} 14}}%
\put(17250,500){\makebox(0,0){\strut{} 16}}%
\put(9625,10050){\makebox(0,0){\strut{}$R$(t)}}%
\put(15200,8750){\makebox(0,0)[r]{\strut{}$R$}}%
\end{picture}%
\endgroup
 

%% file: plot_full_Rc211.tex
\begin{picture}(0,0)%
\includegraphics{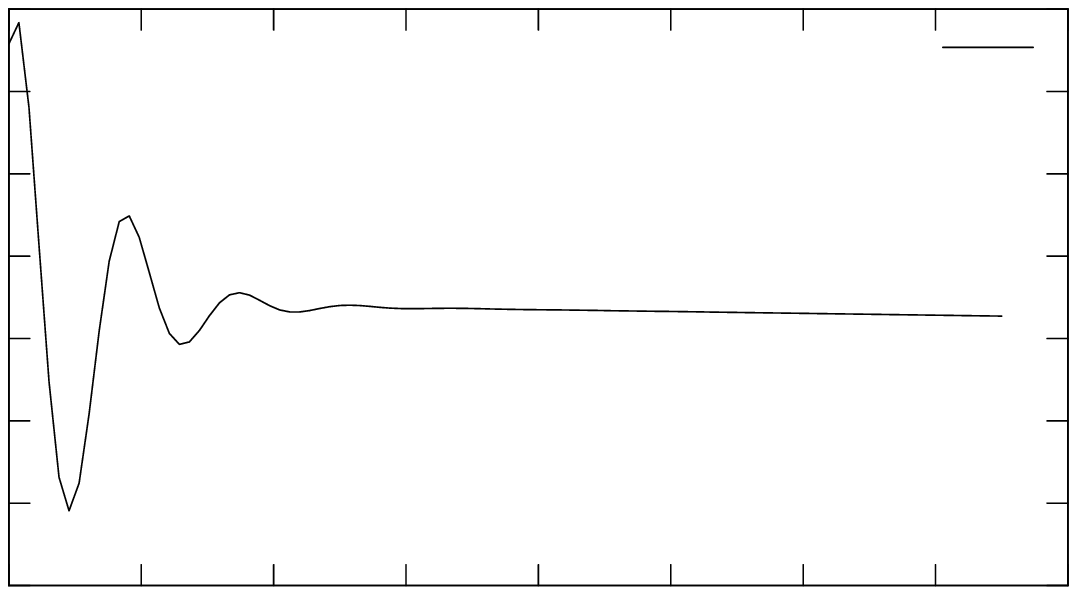}%
\end{picture}%
\begingroup
\setlength{\unitlength}{0.0200bp}%
\begin{picture}(18000,10800)(0,0)%
\put(1750,1000){\makebox(0,0)[r]{\strut{} 0.7}}%
\put(1750,2186){\makebox(0,0)[r]{\strut{} 0.72}}%
\put(1750,3371){\makebox(0,0)[r]{\strut{} 0.74}}%
\put(1750,4557){\makebox(0,0)[r]{\strut{} 0.76}}%
\put(1750,5743){\makebox(0,0)[r]{\strut{} 0.78}}%
\put(1750,6929){\makebox(0,0)[r]{\strut{} 0.8}}%
\put(1750,8114){\makebox(0,0)[r]{\strut{} 0.82}}%
\put(1750,9300){\makebox(0,0)[r]{\strut{} 0.84}}%
\put(2000,500){\makebox(0,0){\strut{} 0}}%
\put(3906,500){\makebox(0,0){\strut{} 2}}%
\put(5813,500){\makebox(0,0){\strut{} 4}}%
\put(7719,500){\makebox(0,0){\strut{} 6}}%
\put(9625,500){\makebox(0,0){\strut{} 8}}%
\put(11531,500){\makebox(0,0){\strut{} 10}}%
\put(13438,500){\makebox(0,0){\strut{} 12}}%
\put(15344,500){\makebox(0,0){\strut{} 14}}%
\put(17250,500){\makebox(0,0){\strut{} 16}}%
\put(9625,10050){\makebox(0,0){\strut{}$R_{ab}R^{ab}$(t)}}%
\put(15200,8750){\makebox(0,0)[r]{\strut{}$R_{ab}R^{ab}$}}%
\end{picture}%
\endgroup
 

%% file: plot_full_0.tex
\begin{picture}(0,0)%
\includegraphics{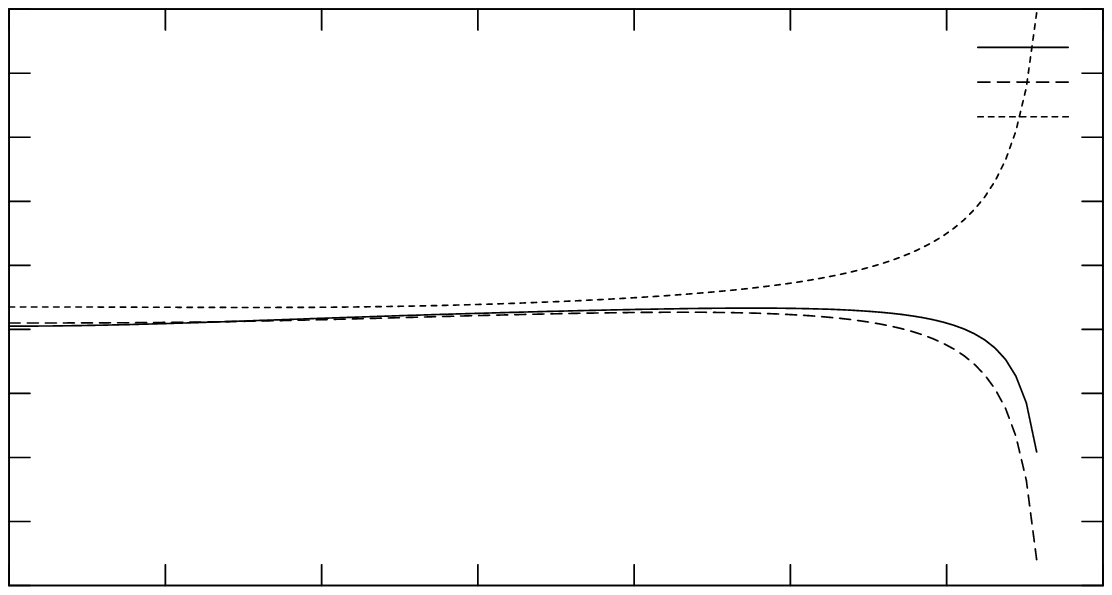}%
\end{picture}%
\begingroup
\setlength{\unitlength}{0.0200bp}%
\begin{picture}(18000,10800)(0,0)%
\put(1250,1000){\makebox(0,0)[r]{\strut{}-8}}%
\put(1250,1922){\makebox(0,0)[r]{\strut{}-6}}%
\put(1250,2844){\makebox(0,0)[r]{\strut{}-4}}%
\put(1250,3767){\makebox(0,0)[r]{\strut{}-2}}%
\put(1250,4689){\makebox(0,0)[r]{\strut{} 0}}%
\put(1250,5611){\makebox(0,0)[r]{\strut{} 2}}%
\put(1250,6533){\makebox(0,0)[r]{\strut{} 4}}%
\put(1250,7456){\makebox(0,0)[r]{\strut{} 6}}%
\put(1250,8378){\makebox(0,0)[r]{\strut{} 8}}%
\put(1250,9300){\makebox(0,0)[r]{\strut{} 10}}%
\put(1500,500){\makebox(0,0){\strut{} 0}}%
\put(3750,500){\makebox(0,0){\strut{} 0.5}}%
\put(6000,500){\makebox(0,0){\strut{} 1}}%
\put(8250,500){\makebox(0,0){\strut{} 1.5}}%
\put(10500,500){\makebox(0,0){\strut{} 2}}%
\put(12750,500){\makebox(0,0){\strut{} 2.5}}%
\put(15000,500){\makebox(0,0){\strut{} 3}}%
\put(17250,500){\makebox(0,0){\strut{} 3.5}}%
\put(9375,10050){\makebox(0,0){\strut{}$\dot{a}$(t)}}%
\put(15200,8750){\makebox(0,0)[r]{\strut{}$\dot{a}_1$}}%
\put(15200,8250){\makebox(0,0)[r]{\strut{}$\dot{a}_2$}}%
\put(15200,7750){\makebox(0,0)[r]{\strut{}$\dot{a}_3$}}%
\end{picture}%
\endgroup
 

%% file: plot_full_v0.tex
\begin{picture}(0,0)%
\includegraphics{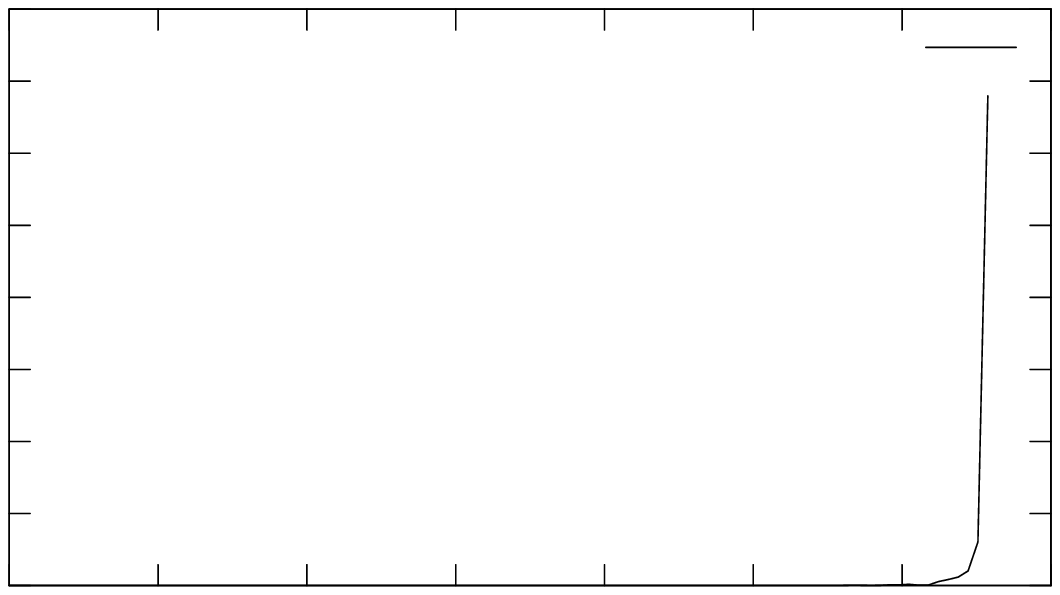}%
\end{picture}%
\begingroup
\setlength{\unitlength}{0.0200bp}%
\begin{picture}(18000,10800)(0,0)%
\put(2000,1000){\makebox(0,0)[r]{\strut{} 0}}%
\put(2000,2037){\makebox(0,0)[r]{\strut{} 1e-11}}%
\put(2000,3075){\makebox(0,0)[r]{\strut{} 2e-11}}%
\put(2000,4112){\makebox(0,0)[r]{\strut{} 3e-11}}%
\put(2000,5150){\makebox(0,0)[r]{\strut{} 4e-11}}%
\put(2000,6187){\makebox(0,0)[r]{\strut{} 5e-11}}%
\put(2000,7225){\makebox(0,0)[r]{\strut{} 6e-11}}%
\put(2000,8263){\makebox(0,0)[r]{\strut{} 7e-11}}%
\put(2000,9300){\makebox(0,0)[r]{\strut{} 8e-11}}%
\put(2250,500){\makebox(0,0){\strut{} 0}}%
\put(4393,500){\makebox(0,0){\strut{} 0.5}}%
\put(6536,500){\makebox(0,0){\strut{} 1}}%
\put(8679,500){\makebox(0,0){\strut{} 1.5}}%
\put(10821,500){\makebox(0,0){\strut{} 2}}%
\put(12964,500){\makebox(0,0){\strut{} 2.5}}%
\put(15107,500){\makebox(0,0){\strut{} 3}}%
\put(17250,500){\makebox(0,0){\strut{} 3.5}}%
\put(9750,10050){\makebox(0,0){\strut{}Constraint}}%
\put(15200,8750){\makebox(0,0)[r]{\strut{}Constraint}}%
\end{picture}%
\endgroup
 

%% file: plot_full_R0.tex
\begin{picture}(0,0)%
\includegraphics{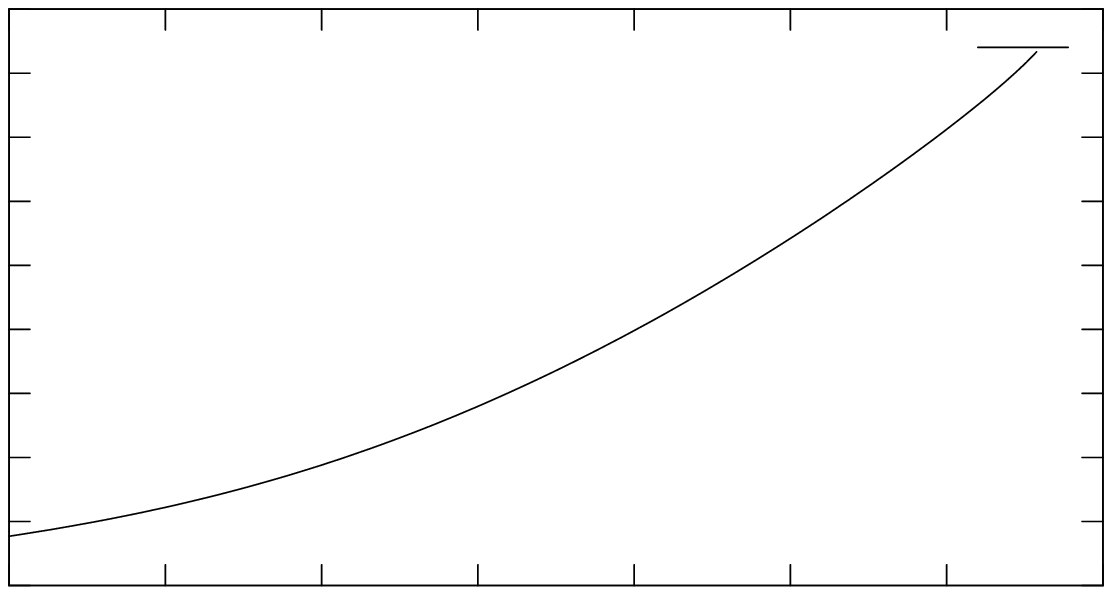}%
\end{picture}%
\begingroup
\setlength{\unitlength}{0.0200bp}%
\begin{picture}(18000,10800)(0,0)%
\put(1250,1000){\makebox(0,0)[r]{\strut{} 0}}%
\put(1250,1922){\makebox(0,0)[r]{\strut{} 2}}%
\put(1250,2844){\makebox(0,0)[r]{\strut{} 4}}%
\put(1250,3767){\makebox(0,0)[r]{\strut{} 6}}%
\put(1250,4689){\makebox(0,0)[r]{\strut{} 8}}%
\put(1250,5611){\makebox(0,0)[r]{\strut{} 10}}%
\put(1250,6533){\makebox(0,0)[r]{\strut{} 12}}%
\put(1250,7456){\makebox(0,0)[r]{\strut{} 14}}%
\put(1250,8378){\makebox(0,0)[r]{\strut{} 16}}%
\put(1250,9300){\makebox(0,0)[r]{\strut{} 18}}%
\put(1500,500){\makebox(0,0){\strut{} 0}}%
\put(3750,500){\makebox(0,0){\strut{} 0.5}}%
\put(6000,500){\makebox(0,0){\strut{} 1}}%
\put(8250,500){\makebox(0,0){\strut{} 1.5}}%
\put(10500,500){\makebox(0,0){\strut{} 2}}%
\put(12750,500){\makebox(0,0){\strut{} 2.5}}%
\put(15000,500){\makebox(0,0){\strut{} 3}}%
\put(17250,500){\makebox(0,0){\strut{} 3.5}}%
\put(9375,10050){\makebox(0,0){\strut{}$R$(t)}}%
\put(15200,8750){\makebox(0,0)[r]{\strut{}$R$}}%
\end{picture}%
\endgroup
 

%% file: plot_full_Rc20.tex
\begin{picture}(0,0)%
\includegraphics{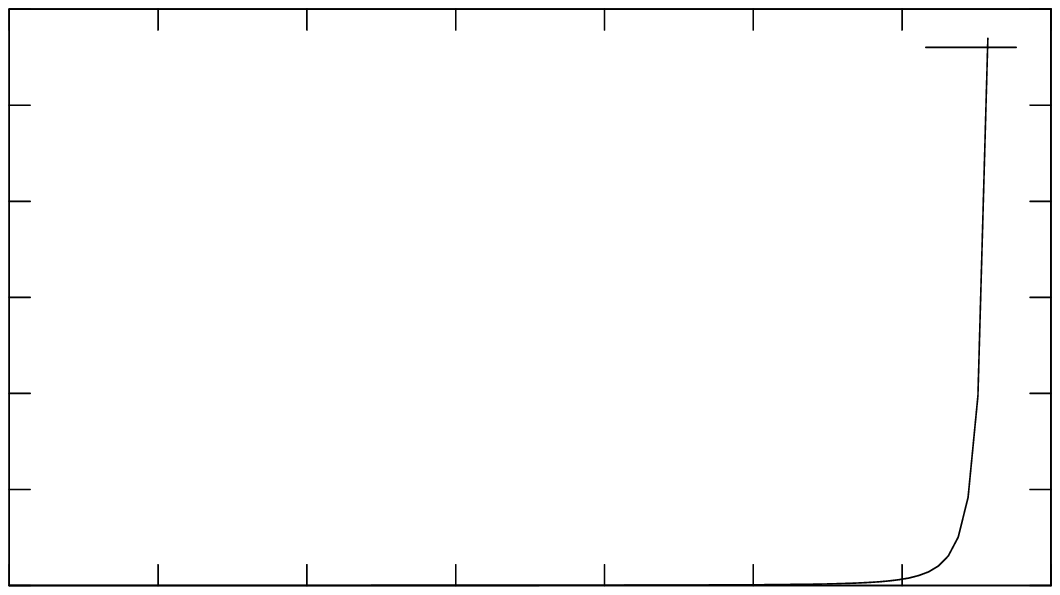}%
\end{picture}%
\begingroup
\setlength{\unitlength}{0.0200bp}%
\begin{picture}(18000,10800)(0,0)%
\put(2000,1000){\makebox(0,0)[r]{\strut{} 0}}%
\put(2000,2383){\makebox(0,0)[r]{\strut{} 5000}}%
\put(2000,3767){\makebox(0,0)[r]{\strut{} 10000}}%
\put(2000,5150){\makebox(0,0)[r]{\strut{} 15000}}%
\put(2000,6533){\makebox(0,0)[r]{\strut{} 20000}}%
\put(2000,7917){\makebox(0,0)[r]{\strut{} 25000}}%
\put(2000,9300){\makebox(0,0)[r]{\strut{} 30000}}%
\put(2250,500){\makebox(0,0){\strut{} 0}}%
\put(4393,500){\makebox(0,0){\strut{} 0.5}}%
\put(6536,500){\makebox(0,0){\strut{} 1}}%
\put(8679,500){\makebox(0,0){\strut{} 1.5}}%
\put(10821,500){\makebox(0,0){\strut{} 2}}%
\put(12964,500){\makebox(0,0){\strut{} 2.5}}%
\put(15107,500){\makebox(0,0){\strut{} 3}}%
\put(17250,500){\makebox(0,0){\strut{} 3.5}}%
\put(9750,10050){\makebox(0,0){\strut{}$R_{ab}R^{ab}$(t)}}%
\put(15200,8750){\makebox(0,0)[r]{\strut{}$R_{ab}R^{ab}$}}%
\end{picture}%
\endgroup
 

%% file: plot_full_4.tex
\begin{picture}(0,0)%
\includegraphics{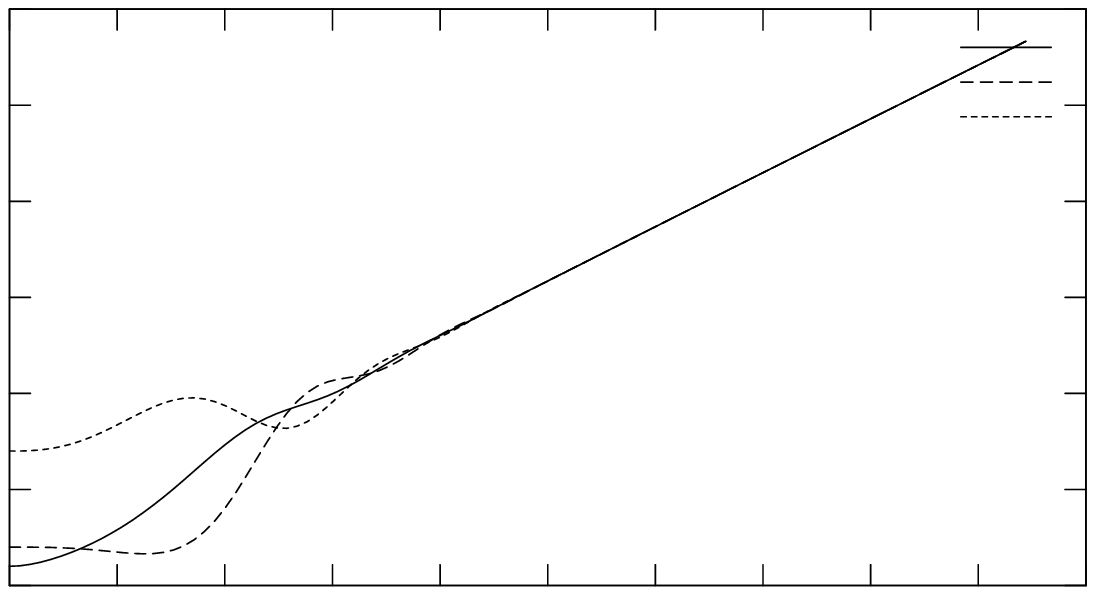}%
\end{picture}%
\begingroup
\setlength{\unitlength}{0.0200bp}%
\begin{picture}(18000,10800)(0,0)%
\put(1500,1000){\makebox(0,0)[r]{\strut{} 0}}%
\put(1500,2383){\makebox(0,0)[r]{\strut{} 0.5}}%
\put(1500,3767){\makebox(0,0)[r]{\strut{} 1}}%
\put(1500,5150){\makebox(0,0)[r]{\strut{} 1.5}}%
\put(1500,6533){\makebox(0,0)[r]{\strut{} 2}}%
\put(1500,7917){\makebox(0,0)[r]{\strut{} 2.5}}%
\put(1500,9300){\makebox(0,0)[r]{\strut{} 3}}%
\put(1750,500){\makebox(0,0){\strut{} 0}}%
\put(3300,500){\makebox(0,0){\strut{} 1}}%
\put(4850,500){\makebox(0,0){\strut{} 2}}%
\put(6400,500){\makebox(0,0){\strut{} 3}}%
\put(7950,500){\makebox(0,0){\strut{} 4}}%
\put(9500,500){\makebox(0,0){\strut{} 5}}%
\put(11050,500){\makebox(0,0){\strut{} 6}}%
\put(12600,500){\makebox(0,0){\strut{} 7}}%
\put(14150,500){\makebox(0,0){\strut{} 8}}%
\put(15700,500){\makebox(0,0){\strut{} 9}}%
\put(17250,500){\makebox(0,0){\strut{} 10}}%
\put(9500,10050){\makebox(0,0){\strut{}$\dot{a}$(t)}}%
\put(15200,8750){\makebox(0,0)[r]{\strut{}$\dot{a}_1$}}%
\put(15200,8250){\makebox(0,0)[r]{\strut{}$\dot{a}_2$}}%
\put(15200,7750){\makebox(0,0)[r]{\strut{}$\dot{a}_3$}}%
\end{picture}%
\endgroup
 

%% file: plot_full_v4.tex
\begin{picture}(0,0)%
\includegraphics{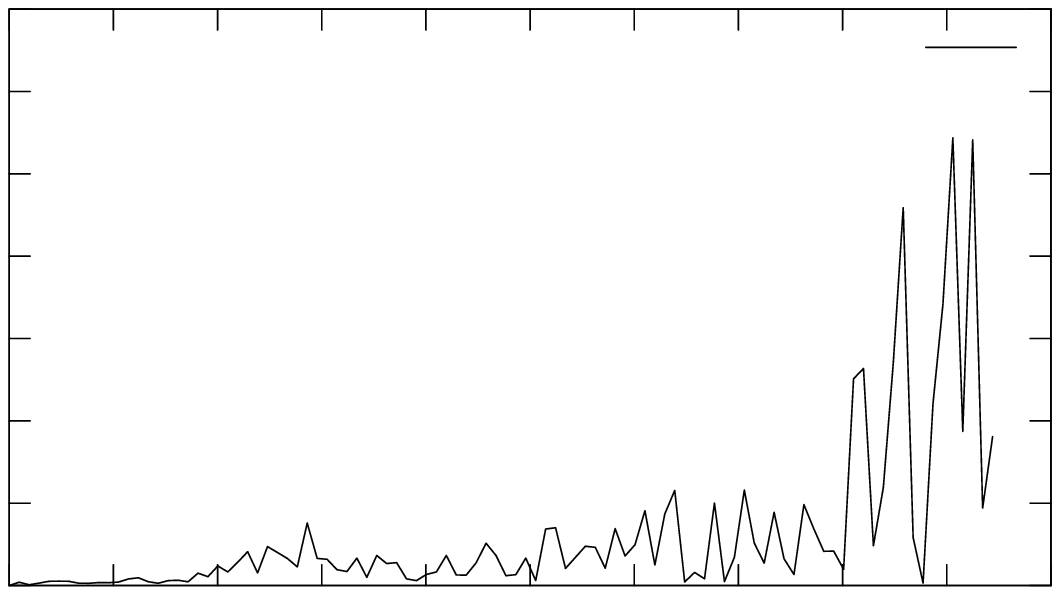}%
\end{picture}%
\begingroup
\setlength{\unitlength}{0.0200bp}%
\begin{picture}(18000,10800)(0,0)%
\put(2000,1000){\makebox(0,0)[r]{\strut{} 0}}%
\put(2000,2186){\makebox(0,0)[r]{\strut{} 1e-14}}%
\put(2000,3371){\makebox(0,0)[r]{\strut{} 2e-14}}%
\put(2000,4557){\makebox(0,0)[r]{\strut{} 3e-14}}%
\put(2000,5743){\makebox(0,0)[r]{\strut{} 4e-14}}%
\put(2000,6929){\makebox(0,0)[r]{\strut{} 5e-14}}%
\put(2000,8114){\makebox(0,0)[r]{\strut{} 6e-14}}%
\put(2000,9300){\makebox(0,0)[r]{\strut{} 7e-14}}%
\put(2250,500){\makebox(0,0){\strut{} 0}}%
\put(3750,500){\makebox(0,0){\strut{} 1}}%
\put(5250,500){\makebox(0,0){\strut{} 2}}%
\put(6750,500){\makebox(0,0){\strut{} 3}}%
\put(8250,500){\makebox(0,0){\strut{} 4}}%
\put(9750,500){\makebox(0,0){\strut{} 5}}%
\put(11250,500){\makebox(0,0){\strut{} 6}}%
\put(12750,500){\makebox(0,0){\strut{} 7}}%
\put(14250,500){\makebox(0,0){\strut{} 8}}%
\put(15750,500){\makebox(0,0){\strut{} 9}}%
\put(17250,500){\makebox(0,0){\strut{} 10}}%
\put(9750,10050){\makebox(0,0){\strut{}Constraint}}%
\put(15200,8750){\makebox(0,0)[r]{\strut{}Constraint}}%
\end{picture}%
\endgroup
 

%% file: plot_full_R4.tex
\begin{picture}(0,0)%
\includegraphics{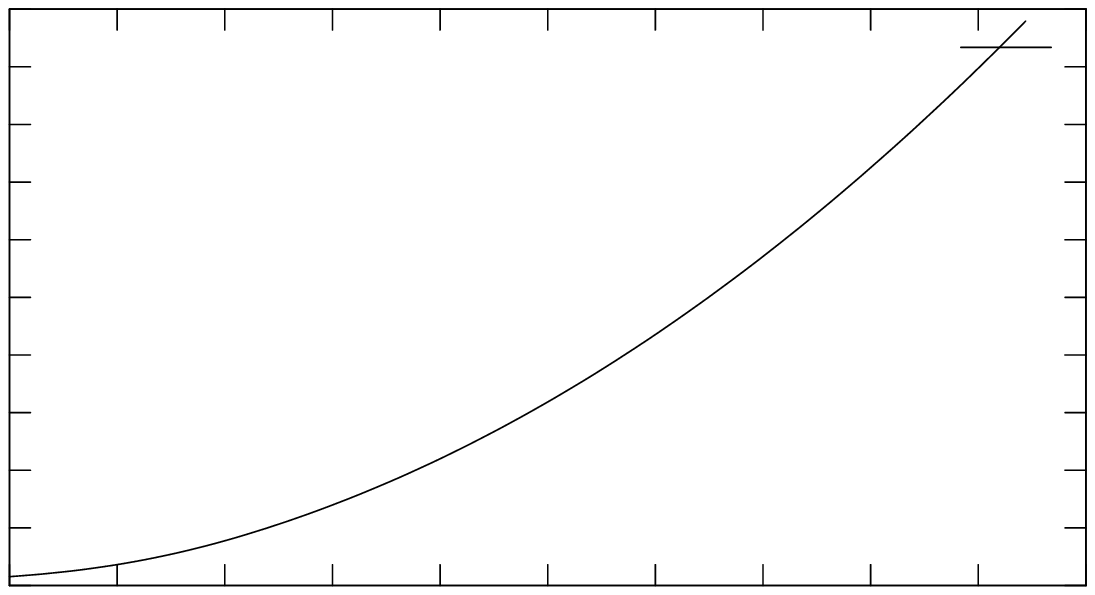}%
\end{picture}%
\begingroup
\setlength{\unitlength}{0.0200bp}%
\begin{picture}(18000,10800)(0,0)%
\put(1500,1000){\makebox(0,0)[r]{\strut{} 0}}%
\put(1500,1830){\makebox(0,0)[r]{\strut{} 10}}%
\put(1500,2660){\makebox(0,0)[r]{\strut{} 20}}%
\put(1500,3490){\makebox(0,0)[r]{\strut{} 30}}%
\put(1500,4320){\makebox(0,0)[r]{\strut{} 40}}%
\put(1500,5150){\makebox(0,0)[r]{\strut{} 50}}%
\put(1500,5980){\makebox(0,0)[r]{\strut{} 60}}%
\put(1500,6810){\makebox(0,0)[r]{\strut{} 70}}%
\put(1500,7640){\makebox(0,0)[r]{\strut{} 80}}%
\put(1500,8470){\makebox(0,0)[r]{\strut{} 90}}%
\put(1500,9300){\makebox(0,0)[r]{\strut{} 100}}%
\put(1750,500){\makebox(0,0){\strut{} 0}}%
\put(3300,500){\makebox(0,0){\strut{} 1}}%
\put(4850,500){\makebox(0,0){\strut{} 2}}%
\put(6400,500){\makebox(0,0){\strut{} 3}}%
\put(7950,500){\makebox(0,0){\strut{} 4}}%
\put(9500,500){\makebox(0,0){\strut{} 5}}%
\put(11050,500){\makebox(0,0){\strut{} 6}}%
\put(12600,500){\makebox(0,0){\strut{} 7}}%
\put(14150,500){\makebox(0,0){\strut{} 8}}%
\put(15700,500){\makebox(0,0){\strut{} 9}}%
\put(17250,500){\makebox(0,0){\strut{} 10}}%
\put(9500,10050){\makebox(0,0){\strut{}$R$(t)}}%
\put(15200,8750){\makebox(0,0)[r]{\strut{}$R$}}%
\end{picture}%
\endgroup
 

%% file: plot_full_Rc24.tex
\begin{picture}(0,0)%
\includegraphics{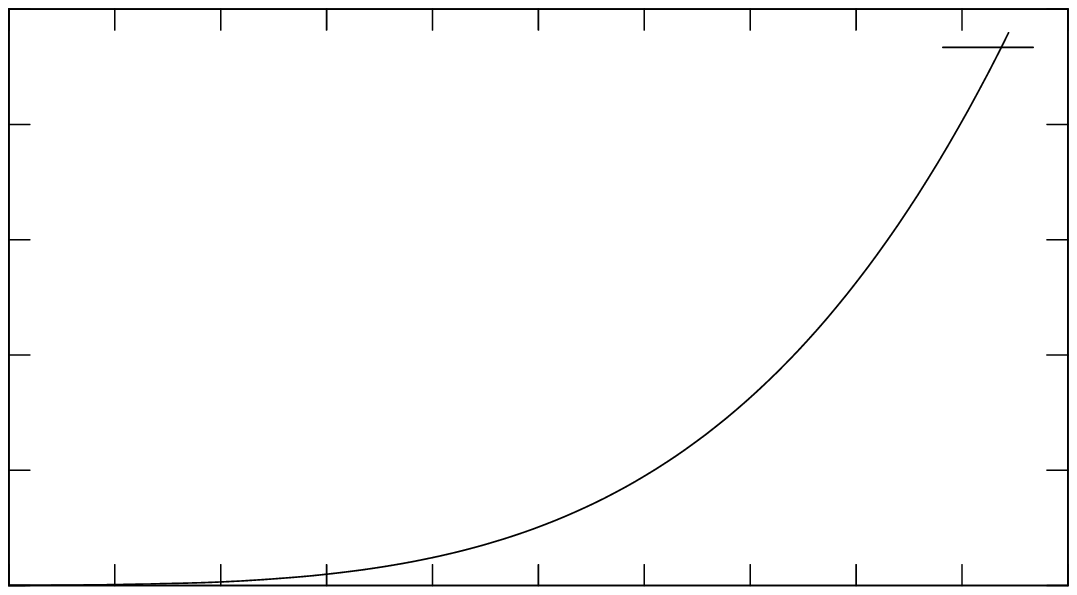}%
\end{picture}%
\begingroup
\setlength{\unitlength}{0.0200bp}%
\begin{picture}(18000,10800)(0,0)%
\put(1750,1000){\makebox(0,0)[r]{\strut{} 0}}%
\put(1750,2660){\makebox(0,0)[r]{\strut{} 500}}%
\put(1750,4320){\makebox(0,0)[r]{\strut{} 1000}}%
\put(1750,5980){\makebox(0,0)[r]{\strut{} 1500}}%
\put(1750,7640){\makebox(0,0)[r]{\strut{} 2000}}%
\put(1750,9300){\makebox(0,0)[r]{\strut{} 2500}}%
\put(2000,500){\makebox(0,0){\strut{} 0}}%
\put(3525,500){\makebox(0,0){\strut{} 1}}%
\put(5050,500){\makebox(0,0){\strut{} 2}}%
\put(6575,500){\makebox(0,0){\strut{} 3}}%
\put(8100,500){\makebox(0,0){\strut{} 4}}%
\put(9625,500){\makebox(0,0){\strut{} 5}}%
\put(11150,500){\makebox(0,0){\strut{} 6}}%
\put(12675,500){\makebox(0,0){\strut{} 7}}%
\put(14200,500){\makebox(0,0){\strut{} 8}}%
\put(15725,500){\makebox(0,0){\strut{} 9}}%
\put(17250,500){\makebox(0,0){\strut{} 10}}%
\put(9625,10050){\makebox(0,0){\strut{}$R_{ab}R^{ab}$(t)}}%
\put(15200,8750){\makebox(0,0)[r]{\strut{}$R_{ab}R^{ab}$}}%
\end{picture}%
\endgroup
 

%% file: plot_full_16.tex
\begin{picture}(0,0)%
\includegraphics{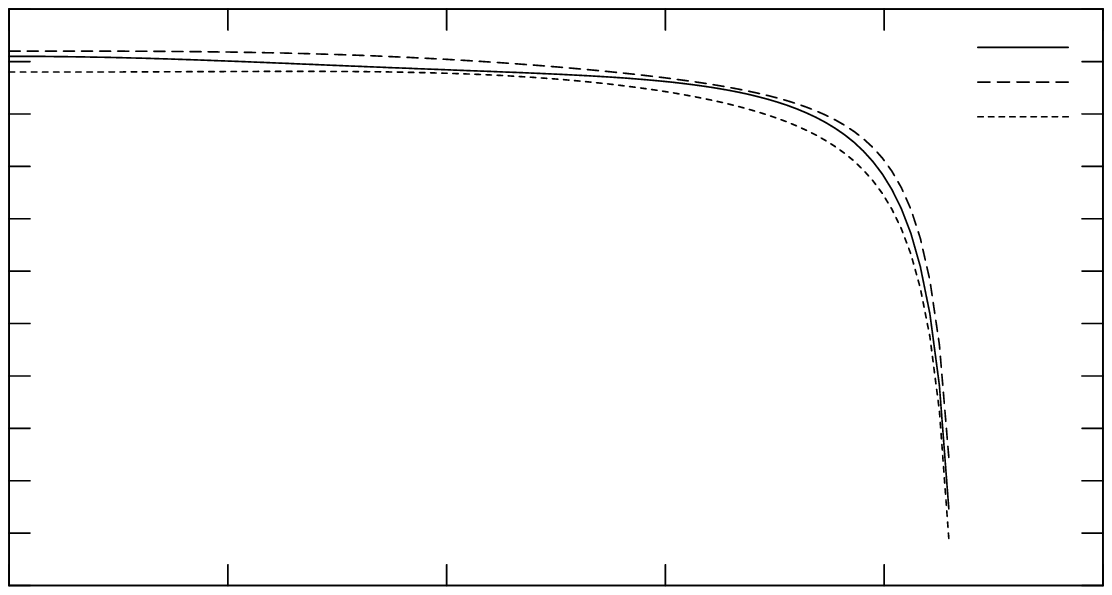}%
\end{picture}%
\begingroup
\setlength{\unitlength}{0.0200bp}%
\begin{picture}(18000,10800)(0,0)%
\put(1250,1000){\makebox(0,0)[r]{\strut{}-10}}%
\put(1250,1755){\makebox(0,0)[r]{\strut{}-9}}%
\put(1250,2509){\makebox(0,0)[r]{\strut{}-8}}%
\put(1250,3264){\makebox(0,0)[r]{\strut{}-7}}%
\put(1250,4018){\makebox(0,0)[r]{\strut{}-6}}%
\put(1250,4773){\makebox(0,0)[r]{\strut{}-5}}%
\put(1250,5527){\makebox(0,0)[r]{\strut{}-4}}%
\put(1250,6282){\makebox(0,0)[r]{\strut{}-3}}%
\put(1250,7036){\makebox(0,0)[r]{\strut{}-2}}%
\put(1250,7791){\makebox(0,0)[r]{\strut{}-1}}%
\put(1250,8545){\makebox(0,0)[r]{\strut{} 0}}%
\put(1250,9300){\makebox(0,0)[r]{\strut{} 1}}%
\put(1500,500){\makebox(0,0){\strut{} 0}}%
\put(4650,500){\makebox(0,0){\strut{} 0.5}}%
\put(7800,500){\makebox(0,0){\strut{} 1}}%
\put(10950,500){\makebox(0,0){\strut{} 1.5}}%
\put(14100,500){\makebox(0,0){\strut{} 2}}%
\put(17250,500){\makebox(0,0){\strut{} 2.5}}%
\put(9375,10050){\makebox(0,0){\strut{}$\dot{a}$(t)}}%
\put(15200,8750){\makebox(0,0)[r]{\strut{}$\dot{a}_1$}}%
\put(15200,8250){\makebox(0,0)[r]{\strut{}$\dot{a}_2$}}%
\put(15200,7750){\makebox(0,0)[r]{\strut{}$\dot{a}_3$}}%
\end{picture}%
\endgroup
 

%% file: plot_full_v16.tex
\begin{picture}(0,0)%
\includegraphics{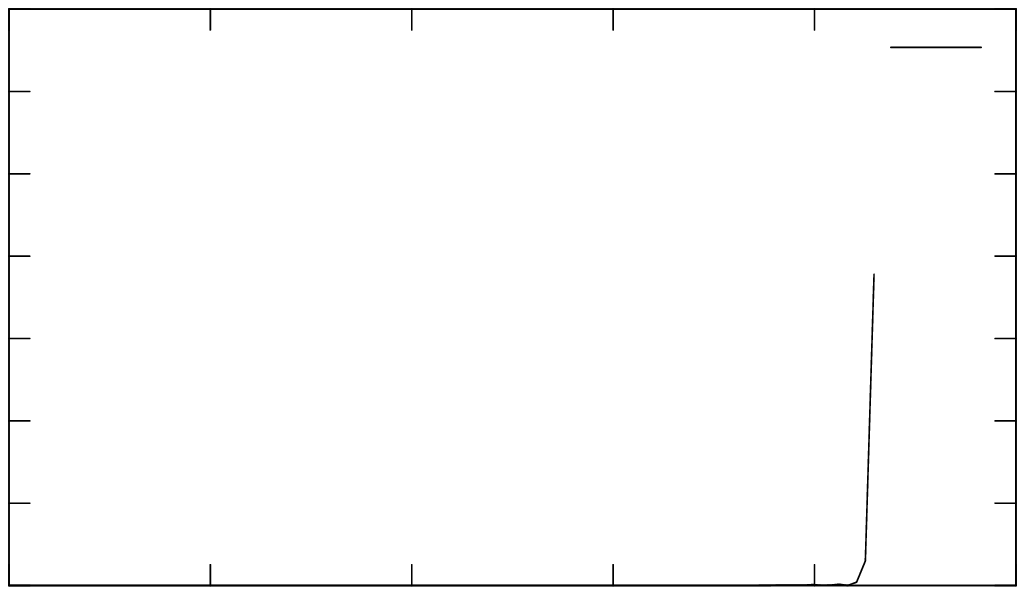}%
\end{picture}%
\begingroup
\setlength{\unitlength}{0.0200bp}%
\begin{picture}(18000,10800)(0,0)%
\put(2500,1000){\makebox(0,0)[r]{\strut{} 0}}%
\put(2500,2186){\makebox(0,0)[r]{\strut{} 2e-10}}%
\put(2500,3371){\makebox(0,0)[r]{\strut{} 4e-10}}%
\put(2500,4557){\makebox(0,0)[r]{\strut{} 6e-10}}%
\put(2500,5743){\makebox(0,0)[r]{\strut{} 8e-10}}%
\put(2500,6929){\makebox(0,0)[r]{\strut{} 1e-09}}%
\put(2500,8114){\makebox(0,0)[r]{\strut{} 1.2e-09}}%
\put(2500,9300){\makebox(0,0)[r]{\strut{} 1.4e-09}}%
\put(2750,500){\makebox(0,0){\strut{} 0}}%
\put(5650,500){\makebox(0,0){\strut{} 0.5}}%
\put(8550,500){\makebox(0,0){\strut{} 1}}%
\put(11450,500){\makebox(0,0){\strut{} 1.5}}%
\put(14350,500){\makebox(0,0){\strut{} 2}}%
\put(17250,500){\makebox(0,0){\strut{} 2.5}}%
\put(10000,10050){\makebox(0,0){\strut{}Constraint}}%
\put(15200,8750){\makebox(0,0)[r]{\strut{}Constraint}}%
\end{picture}%
\endgroup
 

%% file: plot_full_R16.tex
\begin{picture}(0,0)%
\includegraphics{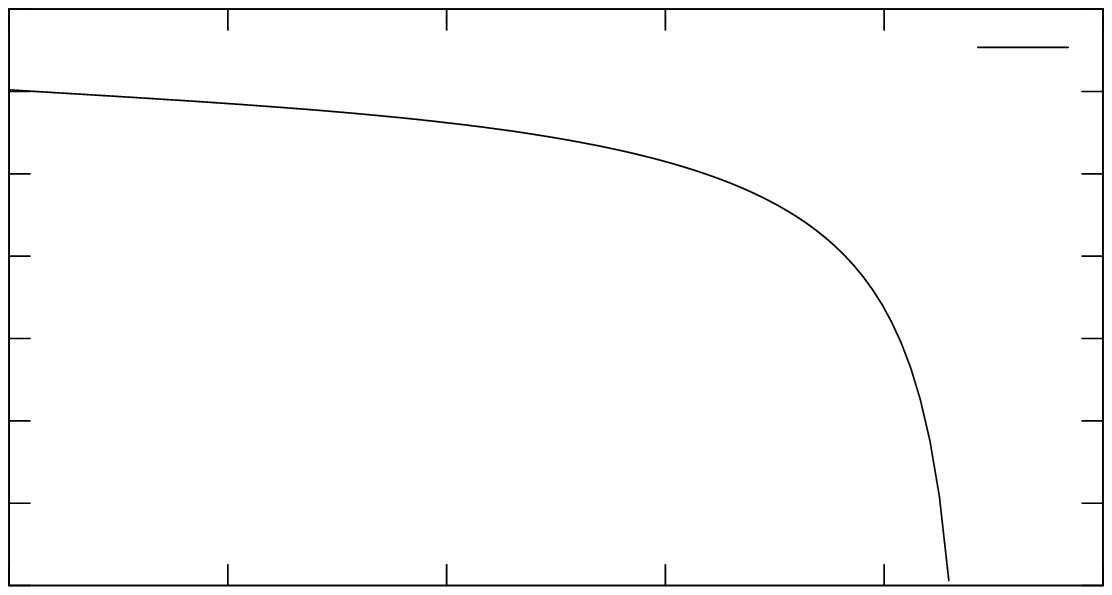}%
\end{picture}%
\begingroup
\setlength{\unitlength}{0.0200bp}%
\begin{picture}(18000,10800)(0,0)%
\put(1250,1000){\makebox(0,0)[r]{\strut{}-30}}%
\put(1250,2186){\makebox(0,0)[r]{\strut{}-25}}%
\put(1250,3371){\makebox(0,0)[r]{\strut{}-20}}%
\put(1250,4557){\makebox(0,0)[r]{\strut{}-15}}%
\put(1250,5743){\makebox(0,0)[r]{\strut{}-10}}%
\put(1250,6929){\makebox(0,0)[r]{\strut{}-5}}%
\put(1250,8114){\makebox(0,0)[r]{\strut{} 0}}%
\put(1250,9300){\makebox(0,0)[r]{\strut{} 5}}%
\put(1500,500){\makebox(0,0){\strut{} 0}}%
\put(4650,500){\makebox(0,0){\strut{} 0.5}}%
\put(7800,500){\makebox(0,0){\strut{} 1}}%
\put(10950,500){\makebox(0,0){\strut{} 1.5}}%
\put(14100,500){\makebox(0,0){\strut{} 2}}%
\put(17250,500){\makebox(0,0){\strut{} 2.5}}%
\put(9375,10050){\makebox(0,0){\strut{}$R$(t)}}%
\put(15200,8750){\makebox(0,0)[r]{\strut{}$R$}}%
\end{picture}%
\endgroup
 

%% file: plot_full_Rc216.tex
\begin{picture}(0,0)%
\includegraphics{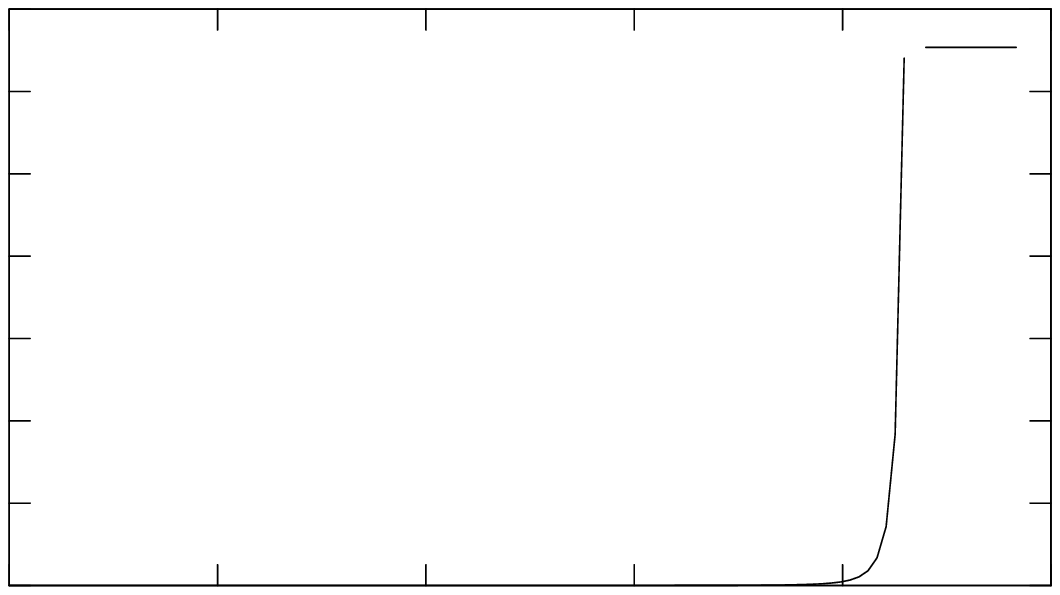}%
\end{picture}%
\begingroup
\setlength{\unitlength}{0.0200bp}%
\begin{picture}(18000,10800)(0,0)%
\put(2000,1000){\makebox(0,0)[r]{\strut{} 0}}%
\put(2000,2186){\makebox(0,0)[r]{\strut{} 10000}}%
\put(2000,3371){\makebox(0,0)[r]{\strut{} 20000}}%
\put(2000,4557){\makebox(0,0)[r]{\strut{} 30000}}%
\put(2000,5743){\makebox(0,0)[r]{\strut{} 40000}}%
\put(2000,6929){\makebox(0,0)[r]{\strut{} 50000}}%
\put(2000,8114){\makebox(0,0)[r]{\strut{} 60000}}%
\put(2000,9300){\makebox(0,0)[r]{\strut{} 70000}}%
\put(2250,500){\makebox(0,0){\strut{} 0}}%
\put(5250,500){\makebox(0,0){\strut{} 0.5}}%
\put(8250,500){\makebox(0,0){\strut{} 1}}%
\put(11250,500){\makebox(0,0){\strut{} 1.5}}%
\put(14250,500){\makebox(0,0){\strut{} 2}}%
\put(17250,500){\makebox(0,0){\strut{} 2.5}}%
\put(9750,10050){\makebox(0,0){\strut{}$R_{ab}R^{ab}$(t)}}%
\put(15200,8750){\makebox(0,0)[r]{\strut{}$R_{ab}R^{ab}$}}%
\end{picture}%
\endgroup
 